\pdfoutput=1
\documentclass{article}
\usepackage{caption}
\usepackage{subcaption}
\usepackage{graphicx}
\usepackage{amsmath}
\usepackage{geometry}
\usepackage{authblk}
\usepackage{hyperref}
\usepackage[T1]{fontenc}
\usepackage[utf8]{inputenc}
\usepackage{float}
\usepackage{csquotes}
\usepackage[version=4]{mhchem}
\usepackage{tikz}
\usepackage[super]{nth}
	
\usetikzlibrary{backgrounds,fit,decorations.pathreplacing}

\title{Quantum Circuit for High Order Perturbation Theory Corrections}
\author[1]{Junxu Li\thanks{Emial: lijunxu1996@gmail.com}}
\author[2]{Xingyu Gao}

\affil[1]{
Department of Physics,
College of Science,

Northeastern University,
Shenyang 110819, China
}
\affil[2]{
Department of Physics and Astronomy,

Purdue University, West Lafayette, IN 47907, United States
}
\begin{document}
\maketitle
	
\begin{abstract}
Perturbation theory (PT) might be one of the most powerful and fruitful tools for both physicists and chemists, which has led to a wide variety of applications.
Over the past decades, advances in quantum computing provide opportunities for alternatives to classical methods.
Recently, a general quantum circuit estimating the low order PT corrections has been proposed.
In this article, we revisit the quantum circuits for PT calculations, and develop the methods for higher order PT corrections of eigenenergy, especially the \nth{3} and \nth{4} order corrections.
We present the feasible quantum circuit to estimate each term in these PT corrections.
There are two the fundamental operations in the proposed circuit.
One approximates the perturbation terms, the other approximates the inverse of unperturbed energy difference.
The proposed method can be generalized to higher order PT corrections.
\end{abstract}

\section*{Introduction}
\label{introduction}

In the recent decades, quantum computing has attracted enormous interests for the potential quantum speedup\cite{preskill2018quantum, cao2019quantum, sajjan2022quantum}.
The scope of its applications have been incrementally broadened into a variety of fields, ranging from prime factorization\cite{shor1994algorithms, hayward2008quantum, lanyon2007experimental, monz2016realization} to data classification\cite{schuld2014quantum, johri2021nearest, li2021quantum, herrmann2022realizing}.
Among these applications, estimating the eigenenergy of many-body systems might be one of the most promising applications of quantum computing\cite{aspuru2005simulated, wang2008quantum, lanyon2010towards, xia2018quantum, mcardle2020quantum, preskill2018quantum, mi2022time, lee2023evaluating}.
In this context, a branch of approaches has been employed to solve the classically intractable electronic structure problems, such as the powerful variational quantum eigensolver\cite{peruzzo2014variational, mcclean2016theory, google2020hartree}, quantum Monte Carlo simulations\cite{foulkes2001quantum, nightingale1998quantum, suzuki1993quantum, bravyi2022quantum}, and the flexible variational quantum simulator\cite{li2017efficient, endo2020variational}.

In our recent work\cite{li2023toward}, we have proposed a general quantum circuit for Perturbation theory (PT) calculations.
Perturbation theory (PT) might be one of the most powerful and fruitful tools for both physicists and chemists, which is a powerful tool to approximate solutions to intricate eigenenergy problems.
In 1926\cite{schrodinger1926undulatory}, Schr\"{o}dinger's proposed the first important application of time-independent PT for quantum systems to obtain quantum eigenenergy.
In our recent work\cite{li2023toward}, we have proposed a general quantum circuit estimating the low order PT corrections: the first order correction for eigenstate, the first and the second order corrections for energy.
The proposed quantum circuit shows potential speedup comparing to the classical PT calculations when estimating second order energy corrections.
Even though, it still remains an open question whether the propose method can be generalized to higher order PT corrections.
In some cases higher order PT corrections are in demand, especially when approximating the accurate energy levels of a complicated system.
One example is M\o ller–Plesset perturbation theory (MPPT)\cite{moller1934note}, which is a typical and powerful PT method in computational chemistry, where the high order corrections are often critical for accurate approximation of molecules.

To address these issues, here we present a general approach to estimate these high order PT corrections of energy, especially for the \nth{3} and \nth{4} order PT corrections of energy.
The proposed method is also feasible for the higher order ones.

Before diving deep, hereafter the time-independent PT method is revisited briefly, which is often termed as Rayleigh–Schr\"odinger PT.
In the time-independent PT method, the Hamiltonian of a complicated unsolved system is described as 
\begin{equation}
    H = H_0 + \lambda V
\end{equation}
$H_0$ represents the unperturbed Hamiltonian, $V$ represents the perturbation and $\lambda\ll 1$.
$H_0$ is often a simple and solvable system, and we have $H_0|\psi_n^{(0)}\rangle = E_n^{(0)}|\psi_n^{(0)}\rangle$, where $|\psi_n^{(0)}\rangle$ and $E_n^{(0)}$ are the eigenstates and corresponding energy levels of $H_0$.
Time-independent PT method leads to the following approximation\cite{griffiths2018introduction}
\begin{equation*}
    E_n = \sum_{k=0}\lambda^kE_n^{(k)}
\end{equation*}
\begin{equation*}
    |\psi_n\rangle = \sum_{k=0}\lambda^k|\psi^{(k)}_n\rangle
\end{equation*}
where $E_n^{(k)}$ indicate the $k$-th order correction of energy, and $|\psi_n^{(k)}\rangle$ indicate the $k$-th order correction of eigenstate.
The $3rd$ order and $4th$ order energy corrections are given as
\begin{equation}
    E_n^{(3)}
    =
    \sum_{k_1}\sum_{k2}
    \frac{V_{nk_2}V_{k_2k_1}V_{k_1n}}{E_{nk_1}E_{nk_2}}
    -V_{nn}
    \sum_{k_2}
    \frac{\left|V_{nk_2}\right|^2}{E^2_{nk_2}}
    \label{eq_e3}
\end{equation}
and
\begin{equation}
    \begin{split}
    E_n^{(4)}
    =&
    \sum_{k_1}\sum_{k_2}\sum_{k_3}
    \frac{V_{nk_3}V_{k_3k_2}V_{k_2k_1}V_{k_1n}}{E_{nk_1}E_{nk_2}E_{nk_3}}
    -
    \sum_{k_1}\sum_{k_3}
    \frac{\left|V_{nk_3}\right|^2}{E^2_{nk_3}}
    \frac{\left|V_{nk_1}\right|^2}{E_{nk_1}}
    \\&
    -2V_{nn}\sum_{k_3}\sum_{k_2}
    \frac{V_{nk_3}V_{k_3k_2}V_{k_2n}}{E^2_{nk_2}E_{nk_3}}
    -V_{nn}^2
    \sum_{k_3}
    \frac{\left|V_{nk_3}\right|^2}{E^3_{nk_3}}
    \label{eq_e4}
    \end{split}
\end{equation}
where all terms involved $k_j$ are summed over $k_j\neq n$, and we introduce the following notations for simplicity
\begin{equation*}
    V_{nm}=\langle \psi_n^{(0)}|V|\psi_m^{(0)}\rangle
\end{equation*}
\begin{equation*}
    E_{nm}=E^{(0)}_n - E^{(0)}_m
\end{equation*}

\section*{Main}
\subsection*{Estimate the \nth{3} order PT correction of energy}
In Eq.(\ref{eq_e3}), there are two summations in the third order PT correction of energy $E^{(3)}_2$.
For simplicity, we denote the first summation in $E^{(3)}_2$ over $k_1,k_2$ as $\epsilon_n^{(3)}$
For the $m^{th}$ order PT correction ($m\geq 2$), the first term $\epsilon_n^{(m)}$ is defined as
\begin{equation}
    \epsilon_n^{(m)}
    =
    \sum_{k_1}\cdots\sum_{k_m}
    V_{nm}
    \left(
    \prod_{j=1}^{m-1}
    \frac{V_{k_{j+1}k_j}}{E_{nk_j}}
    \right)
    V_{k_1n}
    \label{eq_ep}
\end{equation}


\begin{figure}[ht]
    \centering
    \includegraphics[width=1.00\textwidth]{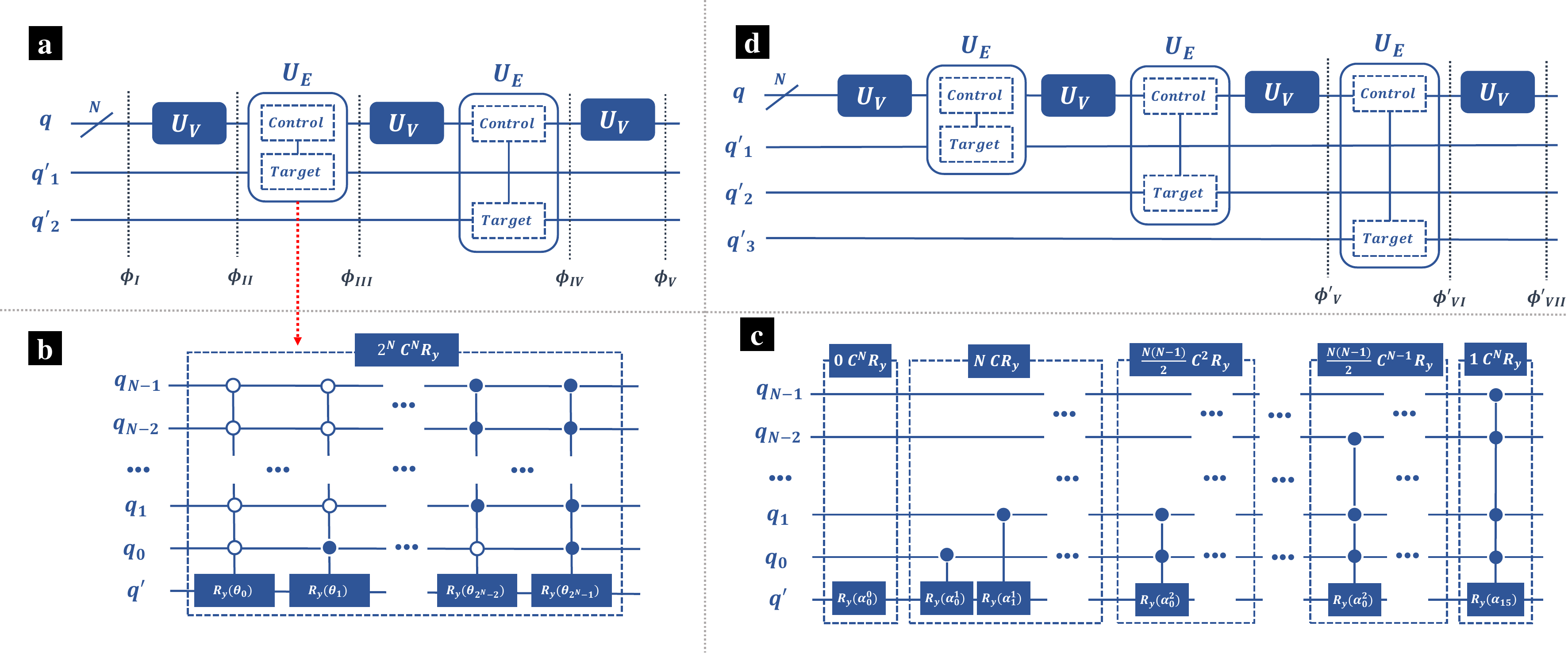}
    \caption{
    {\bf The schematic diagram of quantum circuit estimating $\epsilon_n^{(3)}$ and $\epsilon_n^{(4)}$.}
    (a) The quantum circuit estimating $\epsilon_n^{(3)}$.
    Register $q$ are qubits that represent the given system, whereas $q'$ are ancilla qubits introduced to implement summations.
    $U_V$ is applied on the $q$ qubits to approximate the perturbation $\lambda V$,
    and $U_E$ is applied to prepare the inverse of $E_{nk}$.
    (b) An intuitive decomposition of $U_E$ with $2^N$ $C^NR_y$ gates, where all of the $q$ qubits are control qubits and $q'$ is the target.
    (c) An improved implementation of $U_E$.
    There are $2^N$ multi-controlled $R_y$ gates, but only one of them is $C^NR_y$ gate.
    (d) The quantum circuit estimating $\epsilon_n^{(4)}$.
    }
    \label{fig_main}
\end{figure}

In Fig.(\ref{fig_main}a) a schematic diagram of the quantum circuit estimating $\epsilon_n^{(3)}$ is presented.
Register $q$ are qubits that represent the given system, whereas $q'$ are ancilla qubits introduced to implement summations.
There are in total $N$ $q$ qubits, indicating that there are at maximum $2^N$ energy levels, where it is assumed that there is no degeneracy.
Initially, all $q'$ qubits are prepared at state $|0\rangle$, and $q$ qubits are prepared at state $|n\rangle$, which is the $n^{th}$ state in the computational basis.
Rigorously, state $|n\rangle$ should be written as $|(n)_{bin}\rangle$, where $(n)_{bin}$ is the binary form of $n$ with $N$ digits, and each digit corresponds to a $q$ qubit.
For simplicity, in this article we always use notation $|n\rangle$ instead of $|(n)_{bin}\rangle$. 
In the following discussion, we denote the quantum states at certain steps as $|\phi\rangle$, corresponding to the stages noted by the dashed lines in Fig.(\ref{fig_main}a,d).
After the initialization, we have the overall quantum state $|\phi_{I}\rangle$ as
\begin{equation}
    |\phi_{I}\rangle
    =
    |n\rangle_q\otimes|0\rangle_{q_1'}\otimes|0\rangle_{q'_2}
\end{equation}
where subscripts indicate the corresponding registers.

Next, operation $U_V$ is applied on the $q$ qubits, which approximates the perturbation $\lambda V$.
Generally the perturbation term $\lambda V$ is not unitary, which can not be directly simulated on a quantum computer.
Alternatively, here we approximate the perturbation term $\lambda V$ with $e^{i\lambda V}$, which can be simulated with Trotter decomposition\cite{lloyd1996universal, whitfield2011simulation}.
Thus, $U_V$ can be given as
\begin{equation}
    U_V = T^\dagger e^{i\lambda V} T
    \label{eq_uv1}
\end{equation}
where $T$ is a unitary transformation that converts the eigenstate sets $\{|\psi_k^{(0)}\rangle\}$ into the corresponding computational basis $\{|k\rangle\}$, $T|\psi_k^{(0)}\rangle=|k\rangle$.
More information of $T$ can be found in Methods section.
Eq.(\ref{eq_uv1}) guarantees that 
\begin{equation}
    \langle k_1|U_V|k_2\rangle
    =
    \langle \psi_{k_1}^{(0)}|e^{i\lambda V}|\psi_{k_2}^{(0)}\rangle
    =
    \delta_{k_1k_2} + i\lambda V_{k_1k_2} +\mathcal{O}(\lambda^2)
    \label{eq_uv2}
\end{equation}
After applying $U_V$, the overall quantum state is
\begin{equation}
    |\phi_{II}\rangle
    =
    \left(U_V|n\rangle\right)_q
    \otimes|0\rangle_{q_1'}\otimes|0\rangle_{q'_2}
\end{equation}

We then apply operator $U_E$ to estimate the inverse of $E_{nm}$.
$U_E$ acts on both $q$ and $q'$ qubits, where all $q$ are control qubits and the single $q'$ qubit is target.
$U_E$ is defined as
\begin{equation}
    U_E
    =
    \sum_{k}|k\rangle\langle k|R_y(\theta_k)
\end{equation}
where $\theta$ is defined in Eq.(\ref{eq_theta}).
$U_E$ generates the inverse of $E_{nm}$ with
\begin{equation}
    U_E
    \left(
    |k\rangle_q\otimes|0\rangle_{q'}
    \right)
    =
    \left\{
    \begin{aligned}
        &|k\rangle_q\otimes|0\rangle_{q'},
        &k=n
        \\
        &|k\rangle_q\otimes
        \left(
        \sqrt{1-\frac{C^2}{E_{nk}^2}}|0\rangle
        +\frac{C}{E_{nk}}|1\rangle
        \right)_{q'},
        &k\neq n 
    \end{aligned}
    \right.
    \label{eq_ue}
\end{equation}
where $C$ is a real constant ensuring that $0\leq\left|\frac{C}{E_{nk}}\right|\leq1$.
Intuitively, $U_E$ can be decomposed into $2^N$ $C^NR_y$ gates, where all of the $q$ qubits are control qubits and $q'$ is the target.
In our recent work\cite{li2023toward}, we have presented an improved implementation of $U_E$, which is as depicted in Fig.(\ref{fig_main}c).
In the improved design, there are still $2^N$ multi-controlled $R_y$ gates, but only one of them is $C^NR_y$ gate.
Briefly, there are $\binom{N}{j}$ $C^jR_y$ gates, and the angles $alpha$ are defined as Eq.(\ref{eq_alpha}).
More details about $U_E$ are presented in the Methods section.
The detailed implementation of $U_V$ and $U_E$ can be found in our recent work\cite{li2021quantum, li2023toward}.

$U_E$ converts the overall quantum state as
\begin{equation}
    \begin{split}
        |\phi_{III}\rangle
        =&
        \langle n|U_V|n\rangle|n\rangle_q
        \otimes|0\rangle_{q'_1}\otimes|0\rangle_{q'_2}
        \\
        &+
        \sum_{k_1\neq n}
        \left\{
        |k_1\rangle_q
        \otimes
        \left(
        \langle k_1|U_V|n\rangle
        \sqrt{1-\frac{C^2}{E_{nk_1}^2}}|0\rangle
        +C\frac{\langle k_1|U_V|n\rangle}{E_{nk_1}}|1\rangle
        \right)_{q'_1}
        \otimes|0\rangle_{q'_2}
        \right\}
    \end{split}
\end{equation}
In our recent work\cite{li2021quantum}, we have demonstrated that the first order correction of eigenstate can be obtained from $|\phi_{III}\rangle$, as $V_{mn}$ is approximated by $\langle m|U_V|n\rangle$.
Here our aim is the higher order corrections, and the succeeding operations are still required.
Next, $U_V$ is once again applied on the $q$ qubits, and then $U_E$ is applied on $q$ and $q'_2$, where $q$ qubits are still the control qubits but $q'_2$ is the target.
The new overall quantum state is
\begin{equation}
    \begin{split}
        |\phi_{IV}\rangle
        =&
        \langle n|U_V|n\rangle
        \langle n|U_V|n\rangle
        |n\rangle_{q}\otimes|0\rangle_{q'_1}\otimes|0\rangle_{q'_2}
        \\
        &+
        \langle n|U_V|n\rangle
        \sum_{k_2\neq n}\langle k_2|U_V|n\rangle
        |k_2\rangle_q
        \otimes|0\rangle_{q'_1}\otimes
        \left(
        \langle k_2|U_V|n\rangle
        \sqrt{1-\frac{C^2}{E_{nk_2}^2}}|0\rangle
        +C\frac{\langle k_2|U_V|n\rangle}{E_{nk_2}}|1\rangle
        \right)_{q'_2}
        \\
        &+
        \langle n|U_V|n\rangle
        \sum_{k_1\neq n}
        \left\{
        |n\rangle_q
        \otimes
        \left(
        \langle k_1|U_V|n\rangle
        \sqrt{1-\frac{C^2}{E_{nk_1}^2}}|0\rangle
        +C\frac{\langle k_1|U_V|n\rangle}{E_{nk_1}}|1\rangle
        \right)_{q'_1}
        \otimes|0\rangle_{q'_2}
        \right\}
        \\
        &+
        \sum_{k_1\neq n}
        \sum_{k_2\neq n}
        \left\{
        |k_2\rangle_q
        \otimes
        \left(
        \langle k_1|U_V|n\rangle
        \sqrt{1-\frac{C^2}{E_{nk_1}^2}}|0\rangle
        +C\frac{\langle k_1|U_V|n\rangle}{E_{nk_1}}|1\rangle
        \right)_{q'_1}
        \right.
        \\
        &\qquad\qquad\qquad
        \left.
        \otimes
        \left(
        \langle k_2|U_V|k_1\rangle
        \sqrt{1-\frac{C^2}{E_{nk_2}^2}}|0\rangle
        +C\frac{\langle k_2|U_V|k_1\rangle}{E_{nk_2}}|1\rangle
        \right)_{q'_2}
        \right\}
    \end{split}
\end{equation}

Afterwards, $U_V$ is applied on $q$ qubits for the third time, and we have the overall quantum circuit is
\begin{equation}
    |\phi_{V}\rangle = (U_V)_{q}|\phi_{IV}\rangle
\end{equation}
where the subscript of $U_V$ indicates that it acts on $q$ qubits.
By the end, all qubits are measured.
Theoretically, the probability to get result $|n\rangle_{q}\otimes|1\rangle_{q'_1}\otimes|1\rangle_{q'_2}$ is
\begin{equation}
    \begin{split}
        Pr(n,1,1)
        =&
        \left|
        \left(
        \langle n|_{q}\otimes\langle1|_{q'_1}\otimes\langle1|_{q'_2}
        \right)
        |(U_V)_{q}|\phi_{IV}\rangle
        \right|^2
        \\
        =&
        \left|
        \sum_{k_1\neq n}
        \sum_{k_2\neq n}
        C^2\frac{\langle n|U_V|k_2\rangle
        \langle k_2|U_V|k_1\rangle
        \langle k_1|U_V|n\rangle}
        {E_{nk_1}E_{n_k2}}
        \right|^2
    \end{split}
\end{equation}
By this mean $\epsilon_n^{(3)}$, the first term of $E_n^{(3)}$, can be estimated.


\begin{figure}[ht]
    \centering
    \includegraphics[width=0.45\textwidth]{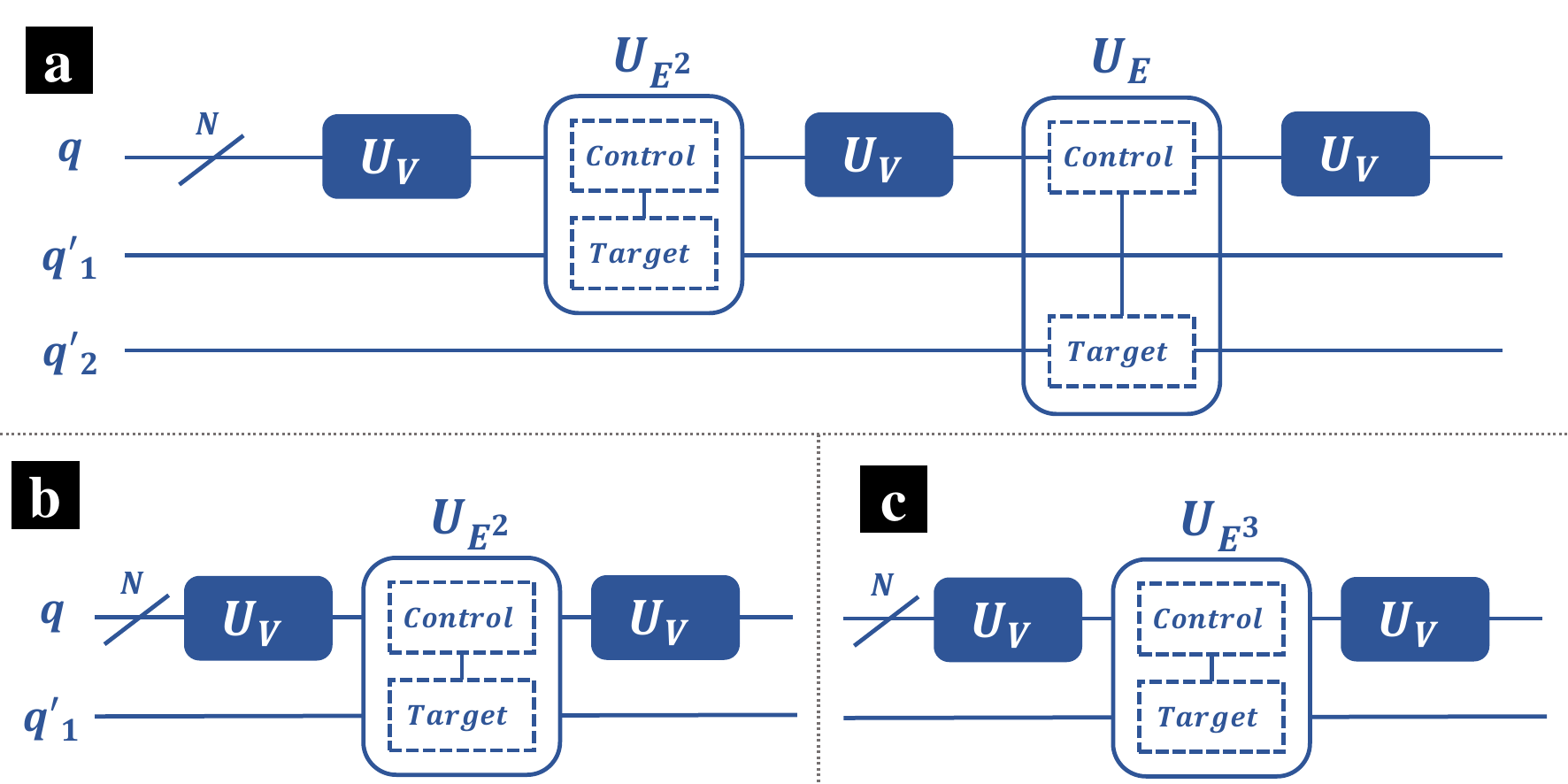}
    \caption{
    {\bf The quantum circuits estimating the miscellaneous terms in $E_n^{(3)}$ and $E_n^{(4)}$.}
    (a) The quantum circuit estimating $\sum_{k_3}\sum_{k_2}
    {V_{nk_3}V_{k_3k_2}V_{k_2n}}/{E^2_{nk_2}E_{nk_3}}$, which appears in the third term of $E_n^{(4)}$, as shown in Eq.(\ref{eq_e4}). 
    (b) The quantum circuit estimating $\sum_{k_2}|V_{nk_2}|^2/E^2_{nk_2}$, which is the second term in $E_n^{(3)}$, as shown in Eq.(\ref{eq_e3}).
    (c) The quantum circuit estimating $\sum_{k_2}|V_{nk_3}|^2/E^3_{nk_3}$, which is the last term in $E_n^{(4)}$, as shown in Eq.(\ref{eq_e4}).
    }
    \label{fig_mis}
\end{figure}

In the second term of $E_n^{(3)}$, $V_{nn}$ is already obtained as the first order correction of energy.
Moreover, $\sum_{k_2}|V_{nk_2}|^2/E^2_{nk_2}$ can be estimated by the quantum circuit as shown in Fig.(\ref{fig_mis}b), where the $U_{E^2}$ is the operation estimating the inverse of $E^2_{nm}$.

\subsection*{Estimate the \nth{4} order PT correction of energy}
There are four terms in the the \nth{4} order PT correction $E_n^{(4)}$.
For simplicity, we denote the first summation in $E^{(4)}_n$ over $k_1,k_2,k_3$ as $\epsilon_n^{(4)}$.
Similarly to $\epsilon_n^{(3)}$, $\epsilon_n^{(4)}$ can be estimated by the quantum circuit as shown in Fig.(\ref{fig_main}d), where $U_E$ and $U_V$ are still harnessed to approximate the perturbations and the inverse of $E_{nk}$.
In addition to the $q$ qubits and the ancilla qubits $q'_1$, $q'_2$, there is one new ancilla qubit noted as $q'_3$.
To avoid confusions, here we denote the overall quantum states in Fig.(\ref{fig_main}d) at certain stage as $|\phi'\rangle$.
At beginning, all $q$ qubits are initialized at state $|n\rangle$, whereas all of the ancilla qubits are prepared at the ground state $|0\rangle$.
The circuit before $|\phi'_V\rangle$ is exactly the same quantum circuit that estimates $\epsilon_n^{(3)}$, as depicted in Fig.(\ref{fig_main}a).
Therefore, we have
\begin{equation}
    |\phi'_V\rangle
    =
    |\phi_V\rangle_{q,q'_1,q'_2}\otimes|0\rangle_{q'_3}
\end{equation}
Next, $U_E$ is applied on $q$ and $q'_3$, where $q$ are the control qubits and $q'_3$ is the target.
The overall quantum state is
\begin{equation}
    \begin{split}
        |\phi'_{VI}\rangle
        =&
        \sum_{k_1\neq n}
        \sum_{k_2\neq n}
        \sum_{k_3\neq n}
        \left\{
        |k_3\rangle_q
        \otimes
        \left(
        \langle k_1|U_V|n\rangle
        \sqrt{1-\frac{C^2}{E_{nk_1}^2}}|0\rangle
        +C\frac{\langle k_1|U_V|n\rangle}{E_{nk_1}}|1\rangle
        \right)_{q'_1}
        \right.
        \\
        &\qquad\qquad\qquad
        \left.
        \otimes
        \left(
        \langle k_2|U_V|k_1\rangle
        \sqrt{1-\frac{C^2}{E_{nk_2}^2}}|0\rangle
        +C\frac{\langle k_2|U_V|k_1\rangle}{E_{nk_2}}|1\rangle
        \right)_{q'_2}
        \right.
        \\
        &\qquad\qquad\qquad
        \left.
        \otimes
        \left(
        \langle k_3|U_V|k_2\rangle
        \sqrt{1-\frac{C^2}{E_{nk_2}^2}}|0\rangle
        +C\frac{\langle k_2|U_V|k_1\rangle}{E_{nk_2}}|1\rangle
        \right)_{q'_3}
        \right\}
        +\cdots
    \end{split}
\end{equation}
where for simplicity, we only present the terms that contribute to the estimation of $\epsilon_n^{(4)}$.
Then $U_E$ is applied on the $q$ qubits, and we have $|\phi'_{VII}\rangle=(U_V)_{q}|\phi'_{VI}\rangle$.
At the end, all of the qubits are measured.
The probability to get result $|n\rangle_{q}\otimes|1\rangle_{q'_1}\otimes|1\rangle_{q'_2}\otimes|1\rangle_{q'_3}$ is
\begin{equation}
    \begin{split}
        Pr'(n,1,1,1)
        =&
        \left|
        \left(
        \langle n|_{q}\otimes\langle1|_{q'_1}\otimes\langle1|_{q'_2}\otimes\langle1|_{q'_3}
        \right)
        |(U_V)_{q}|\phi'_{VI}\rangle
        \right|^2
        \\
        =&
        \left|
        \sum_{k_1\neq n}
        \sum_{k_2\neq n}
        \sum_{k_3\neq n}
        C^3\frac{\langle n|U_V|k_3\rangle
        \langle k_3|U_V|k_2\rangle
        \langle k_2|U_V|k_1\rangle
        \langle k_1|U_V|n\rangle}
        {E_{nk_1}E_{nk_2}E_{nk_3}}
        \right|^2
    \end{split}
\end{equation}

Then consider the second term in $E_n^{(4)}$.
Notice that the summations over $k_3$ and $k_1$ are separable.
$\sum_{k_3}|V_{nk_3}|^2/E^2_{nk_3}$ is already estimated in $E_n^{(3)}$, and $\sum_{k_1}|V_{nk_1}|^2/E^2_{nk_1}$ is the second order correction $E_n^{(2)}$.
Thus, the second term in $E_n^{(4)}$ is known as long as the lower order corrections have been obtained.
The third term in $E_n^{(4)}$, can be estimated by the quantum circuit as shown in Fig.(\ref{fig_mis}a), where $U_{E^2}$ estimates the inverse of $E^2_{nk}$.
As for the last term of $E_n^{(4)}$, $V_{nn}$ is already known, and $\sum_{k_2}|V_{nk_2}|^2/E^3_{nk_2}$ can be estimated by the quantum circuit as shown in Fig.(\ref{fig_mis}c), where the $U_{E^3}$ is the operation estimating the inverse of $E^3_{nm}$, which estimates the inverse of $E^3_{nk}$.

\section*{Discussion}
In this section we will discuss the time complexity of the proposed quantum circuits.
For simplicity, we denote $M=2^N$ as the total number of unperturbed energy levels, and there are in total $N$ $q$ qubits.
Recalling the quantum circuits estimating $\epsilon^{(3)}$ as depicted in Fig.(\ref{fig_main}a), $U_E$ is applied twice and $U_V$ is applied three times.
In the quantum circuits estimating $\epsilon^{(4)}$ as depicted in Fig.(\ref{fig_main}d), $U_E$ is applied three times and $U_V$ is applied four times.
$U_V$ acts on the $q$ qubits, whereas $U_E$ acts on both the $q$ and $q'$ qubits.
Generally, the multi controller gates in $U_E$ consumes more time than the Trotter decomposition in $U_V$.
A $C^NR_y$ gate can be decomposed into $\mathcal{O}(N^2)$ basic gates\cite{barenco1995elementary}.
On the other hand, to prepare a single term $V_{mn}$ with Trotter decomposition, no more than $\mathcal{O}(N)$ basic gates are necessary\cite{whitfield2011simulation}. 
Thus, $U_E$ often denominates the overall time complexity\cite{li2021quantum, li2023toward}.

In the standard design of $U_E$ as depicted in Fig.(\ref{fig_main}b), there are $M$ $C^NR_y$ gates in total, which can be decomposed into $\mathcal{O}(MN^2)$ basic gates.
The total time complexity of the improved design of $U_E$ as shown in Fig.(\ref{fig_main}c) is\cite{li2023toward},
\begin{equation*} 
    \mathcal{O}
    \left(
    \sum_{j=0}^{N}
    \binom{N}{j}j^2
    \right)
\end{equation*}
Time complexity of $U_{E^2}$ and $U_{E^3}$ is exactly the same to $U_E$.
Similarly, $U_{E^2}$ and $U_{E^3}$ denominate the time complexity of the quantum circuits that estimate the miscellaneous terms as shown in Fig.(\ref{fig_mis}).

\begin{figure}[ht]
    \centering
    \includegraphics[width=0.45\textwidth]{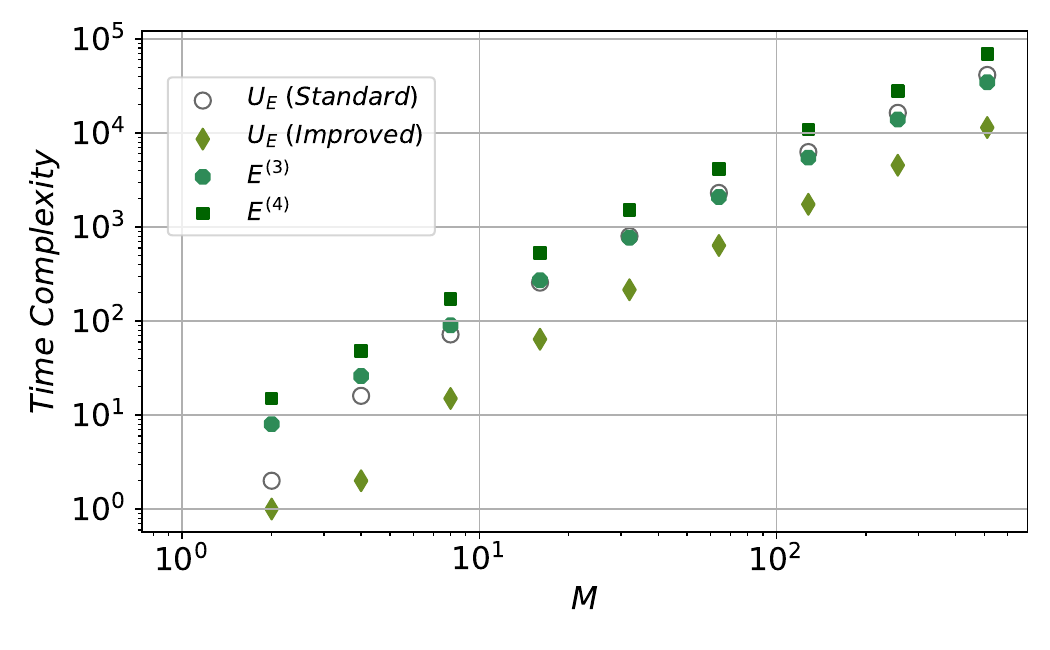}
    \caption{
    {\bf Time complexity of the proposed quantum circuits against the total number of energy levels.}
    Both axes are logarithmic.
    The hollowed circles corresponds to the time complexity of the standard $U_E$ as shown in Fig.(\ref{fig_main}b), whereas the diamonds corresponds to the improved $U_E$ as shown in Fig.(\ref{fig_main}c).
    The octagons and squares indicate the overall time complexity to estimate $E^{(3)}$, $E^{(4)}$, where the improved $U_E$ is applied.
    }
    \label{fig_cmplxty}
\end{figure}

In Fig.(\ref{fig_cmplxty}), we present the overall time complexity of $U_E$, along with the overall time complexity of quantum circuits that estimate $E^{(3)}$, $E^{(4)}$ against $M$, the total number of energy levels.
Time complexity of the standard design of $U_E$, as shown in Fig.(\ref{fig_main}b), is depicted with the hollowed circles, whereas the diamonds corresponds to the time complexity of the improved design of $U_E$ as shown in Fig.(\ref{fig_main}c).
We can find out that the improved $U_E$ consumes much less time than the standard one.
Moreover, we also depicted the overall time consuming to estimate $E^{(3)}$, $E^{(4)}$, corresponding to the octagons and squares, where the improved $U_E$ is applied.
For great $M$, we find out that the time complexity of $U_E$ denominates the overall time consuming to estimate $E^{(3)}$, $E^{(4)}$.


In higher order PT corrections, the $\epsilon_n^{(m)}$ terms can still be described by Eq.(\ref{eq_ep}), where superscript $m$ indicates the $m^{th}$ order correction.
Our approach estimating $\epsilon_n^{(3)}$, $\epsilon_n^{(4)}$ can be generalized to higher order terms $\epsilon_n^{(m)}$, with deeper circuit and more ancilla qubits.
Generally, each $E_{nk}$ in the denominator corresponds to a $U_E$ acting on $q$ and an ancilla digit $q'$.
On the other hand, each $V_{kk'}$ in the numerator corresponds to a $U_V$ acting on the $q$ qubits.
Thus, to estimate $\epsilon_n^{(m)}$ in the $m^{th}$ order correction, $m-1$ ancilla qubits are necessary, and $U_E$ will be applied on the $q$ qubits for $m$ times, whereas $U_E$ will be repeat for $m-1$ times.  


\section*{Conclusions}
In this article, we revisit the quantum circuits for PT calculations, and propose a general quantum circuit to estimate the higher order PT corrections of eigenenergy, especially the \nth{3} and \nth{4} order corrections.
To estimate the PT energy corrections on a quantum computer, the fundamental task is to approximate $V_{nk}$, which describes the perturbation, and $E_{nk}$, which is the difference between unperturbed energy levels.
In this context, $U_V$ and $U_E$ are introduced to estimate these crucial components.
In the proposed quantum circuit, $U_V$ and $U_E$ are applied for several times to estimate the intricate summations in the PT corrections.
We also discuss the time complexity of the proposed quantum circuits.
Generally, $U_E$ denominates the overall time complexity.
Our work smooths the way to implement high order PT calculations, and PT-based methods on with a quantum computer.

\section*{Data availability}
All data that support the plots within this paper and other findings of this study are available from the corresponding author upon reasonable request.

\section*{Acknowledgement}
\label{Acknowledgement}
J.L gratefully acknowledges funding by National Natural Science Foundation of China (NSFC) under Grant No.12305012.

\section*{Author contributions}
J.L. and X.G. discussed the results and wrote the main manuscript text.

\section*{Competing interests}
The authors declare no competing interests.

\section*{Methods}
\subsection*{More details about the transformation $T$}
\label{sec_t}
In PT the unperturbed Hamiltonian $H_0$ is solvable, and thus can be diagonalized as
\begin{equation}
    H_0 = \sum_{n=0}E^{(0)}_n|\psi_n^{(0)}\rangle\langle\psi_n^{(0)}|
\end{equation}
There exists an transformation denoted as $T$, which converts the eigenstate sets $\{|\psi_n^{(0)}\rangle\}$ into the computational basis $\{|n\rangle\}$, $T|\psi_n^{(0)}\rangle=|n\rangle$.
Then we have
\begin{equation}
    T^\dagger H_0T
    =\sum_{n=0}E^{(0)}_n|n\rangle\langle n|
\end{equation}
$T$ is often well-developed for the typical many-body systems.
As an instance, the transformation $T$ that diagonalize the dynamics of strongly correlated quantum systems can be implemented with Bogoliubov transformation and quantum Fourier transformation\cite{verstraete2009quantum}

\subsection*{More details about $U_E$}
\label{sec_ue}
For the standard $U_E$ as depicted in Fig.(\ref{fig_main}b), we have
\begin{equation}
    \theta_{k} = 2\arcsin{
    \left(
    \frac{C}{E_n - E_k}
    \right)
    }
    \label{eq_theta}
\end{equation}
where $C$ is the same constant in Eq.(\ref{eq_ue}).
The improved implementation of $U_E$ can be written as,
\begin{equation}
    U_E=\prod_{x=0}^{M-1}
    \left\{
    \left[
    \bigotimes_{j=1}^{N}
    \left(
    (1-x_j)|0\rangle\langle 0|+|1\rangle\langle 1|
    \right)
    \right]
    \otimes
    R_y(\alpha_x)
    +
    \left[
    I^{\bigotimes N}-
    \bigotimes_{j=1}^{N}
    \left(
    (1-x_j)|0\rangle\langle 0|+|1\rangle\langle 1|
    \right)
    \right]
    \otimes
    I
    \right\}
    \label{eq_ue_improve}
\end{equation}
where $x$ is an integer corresponds to the unperturbed energy level, and $x_j\in\{0,1\}$ is the $j-th$ digit in the corresponding binary form of $x$.
Constraints to the $\alpha$ values can be written as
\begin{equation}
    \sum_{y\in Y(x)}\alpha_y=
    2\arcsin{\left(
    \frac{C}{E_n-E_x}
    \right)}
    \label{eq_alpha}
\end{equation}
where $x$ is the same integer in Eq.(\ref{eq_ue_improve}), and the set $Y(x)$ is
\begin{equation}
    Y(x)=
    \left\{y
    \mid
    y=\sum_{j=0}^{M-1}2^jy_j, y_j\in\{x_j,0\}
    \right\}
    \label{eq_set_y}
\end{equation}
$x_j$, $y_j$ are digits in the binary forms of $x$, $y$.
$U_{E^2}$ and $U_{E^3}$ share the similar structure of $U_E$.
In $U_{E^2}$ and $U_{E^3}$, $E_{nk}$ in Eq.(\ref{eq_theta},\ref{eq_alpha}) are replaced as $E^2_{nk}$ and $E^3_{nk}$.

\bibliography{ref}
\end{document}